\documentclass[a4paper,10pt]{article}

\usepackage[utf8]{inputenc}
\usepackage[english]{babel}
\usepackage{amsmath}
\usepackage{amssymb}
\usepackage{graphicx}
\usepackage{bm}
\usepackage{authblk}
\usepackage{hyperref}
\usepackage{indentfirst}
\usepackage{sidecap}
\usepackage[autostyle, english = american]{csquotes}
\usepackage[left=2cm,right=2cm,top=3cm,bottom=3cm]{geometry}
\usepackage{siunitx}
\MakeOuterQuote{"}

\sidecaptionvpos{figure}{c}

\author[1,2]{K.V.~Nikolaev}
\author[3,4,5,*]{L.I.~Goray}
\author[1,6]{P.S.~Savchenkov}
\author[1]{A.V.~Rogachev}
\author[2]{A.A.~Chouprik}
\author[3]{T.N.~Berezovskaya}
\author[3]{D.V.~Mokhov}
\author[7]{S.A.~Garakhin}
\author[7]{N.I.~Chkhalo}
\author[4,5,8]{A.D.~Buravleuv}
\author[1]{S.N.~Yakunin}

\affil[1]{National Research Center Kurchatov Institute, Moscow, Russia}
\affil[2]{Moscow Institute of Physics and Technology, Dolgoprudny, Russia}

\affil[3]{Alferov University, St Petersburg, Russia}
\affil[4]{Institute for Analytical Instrumentation RAS, St Petersburg, Russia}
\affil[5] {Saint Petersburg Electrotechnical University “LETI”, St. Petersburg, Russia}
\affil[6]{National Research Nuclear University "MEPhI", Moscow, Russia}
\affil[7]{Institute for Physics of Microstructures RAS, Nizhny Novgorod, Russia}
\affil[8]{Ioffe Institute, St. Petersburg, Russia}
\affil[*]{e-mail: \texttt{lig@pcgrate.com}  }

\title{\bf  Conical diffraction of the synchrotron beam to probe the efficiency and morphology of blazed gratings}

\date{July 31, 2025}

\begin{document}

\maketitle

\begin{abstract}

    This study explores the use of synchrotron measurements as a nanometrology tool for blazed gratings. In grazing incidence geometry, one can measure both the conical diffraction and the diffuse scattering on the grating simultaneously in a single scattering pattern.
    The sensitivity of scattering patterns to the structure of the blazed gratings is evaluated.
    The diffraction component of the pattern is shown to be sensitive to the average groove profile of the gratings. 
    Meanwhile, the diffuse scattering depends on the roughness morphology of the reflective surface of blazed gratings.
    These findings are supported by numerical simulations.
    The simulations were performed using several rigorous solvers for the Helmholtz equations, and with a perturbation theory.
    The analysis relies on synchrotron data, as well as data from atomic force microscopy and scanning electron microscopy.
    The aim of this article is to draw the attention of the optical community to the synchrotron measurements as a nanometrology tool for the modern optical elements.

\end{abstract}

\section{Introduction}

    In recent decades, an increasing number of new diffraction-limited sources in the extreme ultraviolet (EUV) and X-ray regions have been built or are in the planning stages. This rapid progress, including the advent of X-ray free-electron lasers (XFELs)~\cite{Huang21}, new-generation synchrotrons~\cite{Chapman23} and compact plasma sources for EUV~\cite{Yang22}, is opening up unprecedented opportunities for scientific research. To realize the full potential of these sources, X-ray optical elements capable of shaping and directing coherent radiation with high precision, preserving wavefront quality and ensuring diffraction-limited resolution in experimental setups play a crucial role \cite{Huang17}. Among the various X-ray optical elements, diffraction gratings are key components. They serve as indispensable tools for spectral analysis and beam monochromatization, especially in the EUV and soft X-ray range, both in grounds and space, where high spectral resolution and high throughput are required.	

    Blazed gratings are particularly important in the so-called "tender X-ray" region ($\sim$1.2--6~keV) for the monochromatization,
    which is a major challenge for conventional optics~\cite{Attwood99}. Multilayer blazed gratings (MLBGs) are highlighted as a promising solution for highly efficient monochromatization element,
    especially in this energy range.
    The study \cite{Sokolov19}, dedicated to the optimization of MLBGs for monochromators, demonstrates the achievement of record efficiencies and the overcoming of the limitations inherent in traditional gratings and crystals. This advantage of blazed gratings in terms of efficiency is further emphasized in high-resolution X-ray spectrometry. In the context of high-resolution spectrometry it was confirmed \cite{Voronov17, Voronov22} that blazed gratings of the new generation (silicon-etched) are the preferred choice for RIXS, providing significantly higher diffraction efficiency compared to lamellar gratings. The development of blazed gratings with high groove density (up to 10,000 lines/mm \cite{Voronov11}) for a next-generation RIXS instrumentation demonstrates the feasibility of creating compact and efficient spectrometric instruments necessary for studying the electronic structure of materials. Moreover, high-frequency MLBGs have a superior efficiency, at least two times, in respect to efficiencies of low-frequency MLBGs and low-frequency bulk gratings in soft X-rays \cite{Voronov15,Voronov16}. For high-frequency MLBGs working in conical (off-plane) diffraction the maximal reached diffraction efficiency can be even more, in higher orders too \cite{Goray16}.
    The efficiency of soft X-ray high-order MLBGs can be increased to nearly the absolute limit of 92\%--98\%.
    
    High-resolution monochromators and spectrometers used in fourth-generation synchrotrons and X-ray free-electron lasers in the soft X-ray and the EUV ranges, as well as numerous planned space missions, require highly efficient and high-power radiation-resistant diffraction gratings of a new generation. It is generally recognized that today such gratings are blazed gratings (with an asymmetric triangular groove profile) produced on vicinal silicon wafers using anisotropic wet etching~\cite{Voronov15, Heilmann24,Goray23,Golub20,Haoyu24}.
    Depending on the lithography method used
    (electron, direct write laser, interference (holographic), photo, ion, nanoimprint, etc.),
    this and accompanying technologies allow the production of gratings with periods $D$ from several tens of nanometers to hundreds of microns and blaze angles from 0.04$^\circ$ to 70$^\circ$.
    The resulting diffraction gratings are characterized by a perfect triangular shape of grooves and a subatomic level of roughness, which allows them to be widely used in devices:
    (1) operating in any, including very high, spectral orders (echelle X-ray gratings \cite{Goray23_2});
    (2) with ultra-high resolution of $\sim 10^{-6}$ (for RIXS~\cite{Voronov15}) and
    (3) with the highest signal-to-noise ratio: $\text{SNR}\sim [10^7, 10^8]$.
    In particular, for such a MLBG with a period of 
    \SI{2}{\micro\metre}
    and a Mo$/$Si multilayer coating,
    we obtained a record absolute efficiency of 40$\%$ at a wavelength of 13.5 nm in the $-8$ order of unpolarized radiation \cite{Goray23_3}.
    
    Among the different types of diffraction gratings, blazed gratings in particular have the ability to concentrate the diffraction radiation into a specific order and therefore offer a significantly higher efficiency compared to gratings with non-triangular groove profiles. Blazed gratings have at least twice the efficiency of lamellar gratings in the short wavelength range; and only this type of gratings can be used for high diffraction orders \cite{Cocco22}. In applications where maximizing photon flux is critical, this property, which results from the concentration of diffracted radiation in a specific order, makes blazed gratings particularly desirable. In order to realize this potential, the grating should have a perfect triangular groove profile and its absorption and scattering should be minimized.
        
    It is also emphasized~\cite{Voronov21} that the development of new synchrotron radiation sources and XFELs, together with progress in EUV lithography, is stimulating the further development of blazed gratings with low and ultra-low blaze angles.
    However, gratings with high and medium blaze angles, which are much easy to fabricate, are required for extreme grazing-incidence conical-diffraction mounts \cite{Seely06,Marlowe16}. Moreover, in opposite to classical (in-plane) diffraction geometry, high-frequency bulk X-ray gratings working in grazing-incidence conical-diffraction with high dispersion and may have any groove dense and, hence, very high spectral resolutions \cite{Petit80,Goray18}.  The planning designs of diffraction limited imaging systems, based on diffraction gratings, opens new horizons for their application in the most demanding fields, including synchrotron and FEL beamline experiments, cold neutron applications, EUV lithography and space missions, where high efficiency and precise control of the spectral characteristics are key factors~\cite{Erko08}. Naturally, the optical design of a grating should be optimized for each spectral range and for each instrumental purpose. In other words, there is a need for continuous refinement and tuning of nanofabrication recipes for blazed gratings. This in turn requires the routine use of metrology, in particular, in solving the inverse diffraction problem. In this regard, a logical approach is to measure and calculate the reflectivity of the grating under working conditions, i.e., with the beam having a specified spectral distribution and a specified incidence-reflection geometry. This provides a definitive metric for a given sample. Another approach to solving more complex grating scatterometry problem is based on separate computations of boundary profiles of multilayer gratings and intensities of short-wave scattering~\cite{Goray13}. On the other hand, sometimes, instead of simply checking whether a sample is suitable or not, it is necessary to investigate the physical properties of the scattering itself on a given sample. For this purpose, synchrotron measurements should be used first of all.
        
    We focus on studying blazed grating scattering with synchrotron radiation. Specifically, we consider measuring the conical diffraction in the grazing incidence small angle X-ray scattering (GISAXS) experiment (see examples with regard to gratings in~\cite{Soltwisch17,Pfluger17,FernandezHerrero22}. In the GISAXS experiment, the sample is irradiated with a monochromatic X-ray beam at a small angle of incidence, close to the region of total external reflection. Under such illumination conditions, almost the entire sample is irradiated in the direction of the beam path. Therefore, the scattering pattern does not correspond to the local structure, but rather to the statistical distribution of the structure parameters across the illuminated spot. Thus, by analyzing the scattering pattern, one can estimate the performance of an entire grating. The scattering pattern itself is measured with a 2D detector, so it can be conveniently recorded in a single exposure without scanning over multiple sample-detector orientations. To fully utilize the convenience of the 2D detector, the sample is oriented for conical diffraction. In conical diffraction, the grating grooves are oriented along the path of the beam. The diffraction peaks then appear on the circle formed by the cross section of the detector surface and the cone, with the angle equal to the angle of incidence. The relevant information, i.e. the intensity of the different diffraction peaks, is thus distributed both vertically and horizontally on the detector. This is in contrast to classical diffraction geometry, where the grooves are perpendicular to the path and the diffraction peaks are distributed on a vertical line on the detector.
    
    However, for samples with a sufficiently large period to probe wavelength ratio, it is often problematic in conical diffraction to discriminate between neighboring diffraction peaks. In this article, we show how to get around this problem. We demonstrate this experimentally. As a demonstration sample, we use the blazed grating designed for in-plane EUV applications which has the submicron period of $\sim$ 400 nm and blaze angle of $\sim 4^{\circ}$, and probe it with hard X-rays at subnanometer wavelengths. We support this claim by comparing the GISAXS data with the diffraction simulations. The forward model in the simulations is based on atomic force microscopy characterization of the sample. We cross-validate the mathematical model of diffraction using different solvers. We also show that in addition to the diffraction efficiency, the GISAXS data contain insightful information about the physics of scattering on blazed gratings. We do this by analyzing the distribution of diffuse scattering. Taken together, we argue that synchrotron metrology can prove to be a useful addition to the design and manufacturing cycle of optical elements based on blazed gratings.

\section{Sample design and pre-characterization}
    \begin{SCfigure}[][b!]
        \centering
        \includegraphics{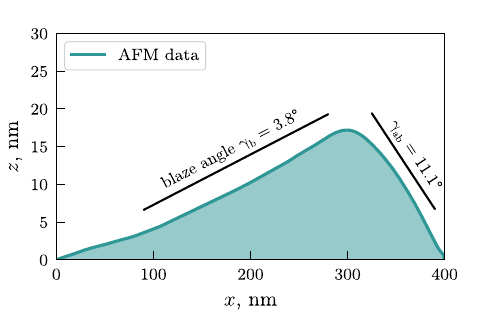}
        \caption{Asymmetric triangular profile of an actual blazed grating sample,
                 as characterized with AFM.
                 The groove profile is taken as an average over multiple periods of the grating.
                 The tangent angles of the blazed and anti-blazed surfaces are marked in the figure.
                 The angles are calculated from the linear fit of the flat part of the profile,
                 and the corresponding lines are shown.
                }
        \label{fig:AFM}
    \end{SCfigure}

    The ideal optical design of a blazed grating have the asymmetrical triangular groove profile.
    The performance of a blazed grating is as good as the real structure is close to this ideal triangular profile with smooth and flat surface of the reflecting faces.
    Thus, there are two reflecting surfaces:
    blaze and anti-blaze surfaces, which are characterized by corresponding tangent angles
    $\gamma_{\rm b}$ and $\gamma_{\rm ab}$, respectively.
    The blaze face of the triangle is the working reflective surface of such a device,
    hence the efficiency of a blazed grating is determined by the quality of this surface,
    i.e. the area and the roughness.
    To optimize these parameters, we select three basic quality criteria.
    First, the length of the reflecting surface $l$ should be more than~75\% of the period:
    $l \ge 0.75D$.
    Second, the asymmetry coefficient $k = \gamma_{\rm ab} / \gamma_{\rm b}$ should satisfy $k \ge 5$.
    Finally, the roughness of the reflecting surface should be minimized.
    We have detailed the considerations of the effect of roughness on the diffraction efficiency of blazed gratings in~\cite{Goray23_2}.

    As part of the refinement of the blaze grating optical design,
    the results of which are described in~\cite{Goray23},
    a series of prototypes was prepared.
    For this study, a prototype was selected as a working sample to demonstrate the capabilities of synchrotron measurements as a metrological tool for blaze gratings.
    The grating has a period of $D = \SI{0.4}{\micro\metre}$;
    it is fabricated on the surface of a polished vicinal silicon wafer.
    The grating is covered with a 25~nm~reflective platinum coating.
    The nominal blaze angle is $\gamma_{\rm b} = 4^\circ$.
    The sample has been manufactured as follows.
    The wafer was prepared so that the miscut angle between its (111) surface and the [112] crystallographic direction would correspond to the defined blaze angle: $\gamma_{\rm b} = 4^\circ$.
    Then the e-beam lithography was used.
    The e-beam exposed a field size of $10\times10$~mm$^2$ on the wafer.
    Then, a triangular structure was created by an anisotropic etching in the potassium hydroxide.

    Further, we used atomic force microscopy (AFM) to pre-characterize the pattern etched onto the silicon wafer's surface.
    This analysis served two purposes: to analyze the shape of the pattern and to estimate the reflective surface roughness's statistical parameters.
    Groove profile analysis involved taking two measurements at different locations on the surface,
    each with a field size of
    $1\times1$~\SI{}{\micro\metre}.
    The profile is estimated as an average within each measurement, and then between measurements.
    The resulting groove profile is shown in Fig.~\ref{fig:AFM}.
    It has the characteristic asymmetric triangular shape.
    The average groove profile of the silicon grating before coating has a blaze angle of $\gamma_{\rm b} =3.6^\circ$ and an anti-blaze angle of
    $\gamma_{\rm ab}= 12.5^\circ$,
    giving an asymmetry parameter of $k = 3.5$.
    The average groove profile of the coated grating has a blaze angle of $\gamma_{\rm b} =3.8^\circ$ and an anti-blaze angle of
    $\gamma_{\rm ab}= 14.7^\circ$,
    giving an asymmetry parameter of $k = 3.9$.
    This AFM analysis shows that the design of this sample is not yet optimal, with a $k$ value of less than 5.
    The $\gamma_{\rm b}$ was estimated using only the flat part of the reflective surface.
    Furthermore, the reflective surface profile is curved rather than flat at the beginning.
    In the following sections, we will use this data to estimate the optical performance of the 
    blaze grating based on this design.
    
\begin{figure}[t!]
        \centering
        \includegraphics{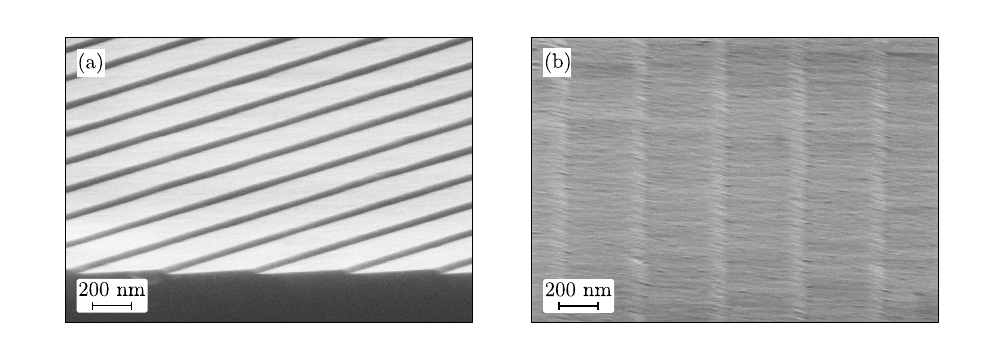}
        \caption{
                    SEM images of the blaze grating sample:
                    (a) Side view;
                    (b) Top view.
                }
        \label{fig:SEM}
    \end{figure}

    We also use AFM data to characterize the roughness of reflective surfaces of the silicon grating.
    To do so, measurements at two different spatial scales were taken.
    Short-scale measurements (\SI{1}{\micro\metre}) were used to estimate the roughness r.m.s. amplitude,
    resulting in $\sigma = 0.5$~nm.
    Long-scale measurements (\SI{20}{\micro\metre}) were performed to estimate the correlation length, yielding an estimate of $\xi_\parallel = 1.5$~nm.
    Next, a homogeneous reflective coating was deposited on the surface of the blazed grating.
    The platinum coating layer was deposited using magnetron evaporation.
    The nominal thickness of the coating is $h = 25$~nm.
    The thickness and uniformity of the coating were controlled using a planar witness sample, followed by X-ray reflectivity analysis.
    Moreover, the uniformity of the coating was inspected using scanning electron microscopy (SEM).
    The resulting SEM images are shown in Fig.~\ref{fig:SEM}.
    Finally, another series of AFM measurements was performed to estimate the roughness statistics of the reflective coatings surface.
    The AFM data were gathered in three spatial ranges:
    \SI{32}{\micro\metre}, \SI{10}{\micro\metre},
    and \SI{2}{\micro\metre}.
    The roughness measured at the \SI{2}{\micro\metre} scale is $\sigma = 0.6$~nm,
    which is consistent with the roughness of the surface before coating.
    Therefore, we can assume that the surface morphology was not drastically modified by the coating.
    Synchrotron measurements were then performed.

\section{Synchrotron measurements}

    Synchrotron measurements were performed at the "Langmuir" bending magnet beamline of the Kurchatov Synchrotron.
    This beamline is primarily designed for soft matter measurements
    with optical design aimed for a parallel beam~\cite{Yakunin22}.
    However, it can also be configured to measure solid-state samples, when exposure with a parallel beam is required, which applies to our case.
    During the measurements the optical configuration included a thermally stabilized,
    two-bounce silicon monochromator with a (111) reflection.
    Higher harmonics of the monochromated beam were suppressed using quartz and tungsten X-ray mirrors.
    The synchrotron beam was collimated with three sets of slits.
    The resulting beam size was $200 \times 100$~\SI{}{\micro\metre}.
    The corresponding average direct-beam intensity is approximately $3 \cdot 10^7$~counts/s.
    Both the vertical and the horizontal beam divergence in this configuration was less then 5~arcsec.
    The measurements were performed at a photon energy of $E = 9$~keV and the scattering pattern was recorded with a Lambda~750K detector.

    \begin{figure}[t!]
        \centering
        \includegraphics{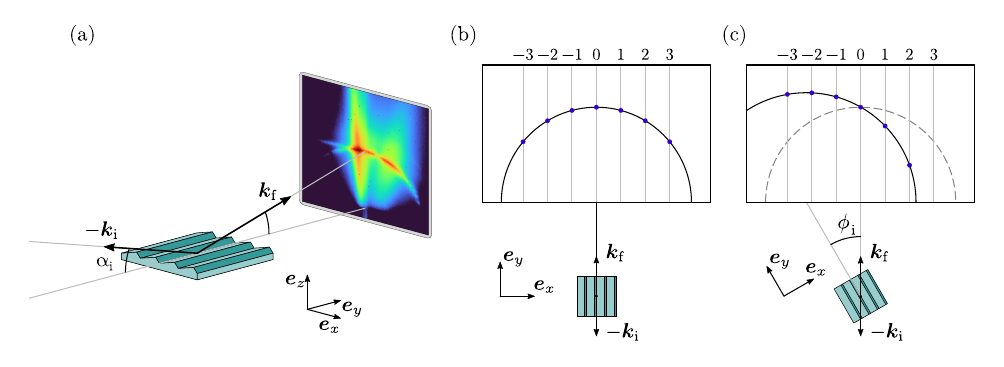}
        \caption{(a) Sketch of the conical diffraction measurement on the blazed grating in the GISAXS experiment.
                 (b) A schematic representation of the positions of the diffraction peaks on the detector in the conical geometry. The numbers represent the order of diffraction $m$. This pattern is the result of the spherical dispersion Eq.~\ref{eq:spherical}.
                 (c) The change of the positions of the diffraction peaks when the grating is rotated in the lateral plane on the angle $\phi_i$.
                 The position of the specularly reflected beam ($m=0$) is unchanged, while the diffraction cone moves with the sample. In this way, the separation between adjacent peaks on the detector can be increased.
                }
        \label{fig:sketch}
    \end{figure}

    The sample-detector geometry for conical diffraction measurements in the GISAXS experiment is shown in the sketch in Fig~\ref{fig:sketch}~(a).
    The sample is illuminated at an angle of incidence of $\alpha_i$ by a monochromatic wave with the wave vector $\bm{k}_i$.
    The grooves of the blazed gratings are oriented such that the projection of the wave vector onto the surface $\bm{k}_{i\parallel}$ is parallel to the grooves of the gratings.
    Diffraction peaks are distributed on the circle formed by the section of the cone with angle $\alpha_i$. Their exact positions are determined by a spherical dispersion
    relation~\cite{Mikulik99}.
    To calculate it, first consider that each pixel on the detector is associated with a particular scattering direction $\bm{k}_f$, and thus corresponds to a point in reciprocal space
    $\bm{q} = -\bm{k}_i+\bm{k}_f$.
    According to Laue's condition,
    the $m$-th diffraction order is at $\bm{q}_\parallel = 2\pi m \hat{\bm{e}}_x/D$, 
    where the basis vector of the reference frame is bound to the sample orientation as shown in Fig.~\ref{fig:sketch}~(b),
    and $D$ is the period of the grating.
    The vertical position of the $m$-th diffraction order on the detector is then derived by considering that the scattering is elastic:
    \begin{equation}
        q_z = \sqrt{
            k^2-(\bm{k}_{i\parallel}+2\pi m \hat{\bm{e}}_x/D)^2
        } - k_{iz}
        ,
        \label{eq:spherical}
    \end{equation}
    where $k = 2\pi/\lambda$ is the wavenumber in the ambient and $\lambda$ is the wavelength in vacuum.
    This yields a generalized version of the grating equation which accounts for the conical diffraction:
    \begin{equation}
        -m \lambda / \sin \alpha_i = D (\sin \phi_i - \sin \phi_f),
        \label{eq:grating_equation}
    \end{equation}
    where $\phi_i$ and $\phi_f$ are the angles of rotation in the dispersive plane corresponding to the incidence onto the surface and to the diffraction, respectively,
    and $\alpha_i$ is the grazing incidence angle.
    In terms of experiment on the beamline,
    the $\phi_i$ angle is fully defined by the sample-beam orientation,
    and $\phi_f$ simply encodes the position of a pixel on the detector.
    Using Eq.~\ref{eq:grating_equation},
    one can calculate the direction of the $m$-th diffraction order in conical diffraction.
    The extreme grazing-incidence conical diffraction mounting in which the direction of incident radiation is confined to a plane parallel to the direction of the grooves has the unique property of maintaining a maximal level of the diffraction efficiency for X-rays \cite{Werner77}. Such an order efficiency may be very close to the reflectance of the respective mirror at any wavelength due to the specific geometry of the incident radiation in respect to blaze grooves \cite{Petit80}.
    
    The geometric interpretation is straightforward.
    Similar to diffraction in other laterally periodic systems~\cite{Robinson86,Petukhov03}, the diffraction intensity is distributed along Bragg rods
    [vertical lines in Fig.~\ref{fig:sketch}~(b)], and for ideal periodic systems the conditions are satisfied at the intersection with the diffraction cone.
    The $0$-th order corresponds to the specular reflection.
    Its position is entirely defined by the incident beam and the surface plane,
    and the grating structure on the surface does not affect it.
    It becomes apparent if the grating is adjusted away from the conic orientation.
    In Fig.~\ref{fig:sketch}~(c) the grating is rotated along the normal to the surface on the angle $\phi_i$.
    According to Eq~\ref{eq:spherical}, the position of the specular beam ($m=0$) is unchanged. However, the diffraction cone shifts to maintain its center along the grooves of the gratings.
    The specular beam is no longer at the apex of the diffraction cone and the radius of the cone section is increased to intersect the specular beam.
    
    Another result of Eq.~\ref{eq:spherical} is that as the ratio $D/\lambda$ increases, the positions of the diffraction peaks on the diffraction cone become denser. Therefore, at certain $D/\lambda$, the separation between adjacent peaks becomes smaller than the instrument resolution and the diffraction merges into the continuous circle.
    However, again according to Eq.~\ref{eq:spherical}, by changing the angle $\phi_i$ one can always adjust the scattering geometry so that there is a diffraction peak near $q_z = 0$, while the specular peak is fixed at $q_z = -2k_{iz}$. Thus, the vertical separation between the diffraction peaks can be increased by adjusting away from the conical orientation at a certain angle $\phi_i$, which can be calculated from Eq.~\ref{eq:spherical}.

    In the experiments, a series of GISAXS patterns were measured at different angles of incidence and for two sample orientations:
    exact conical mounting ($\phi_i = 0^\circ$) and a tilted conical mounting with $\phi_i = 0.3^\circ$.
    Each such GISAXS pattern contains the conical diffraction peaks.
    They also contain features attributed to the diffuse scattering.
    The next section considers these two types of features and the information they provide about the structure.

\section{Results and discussion}

\begin{figure}[t!]
        \centering
        \includegraphics{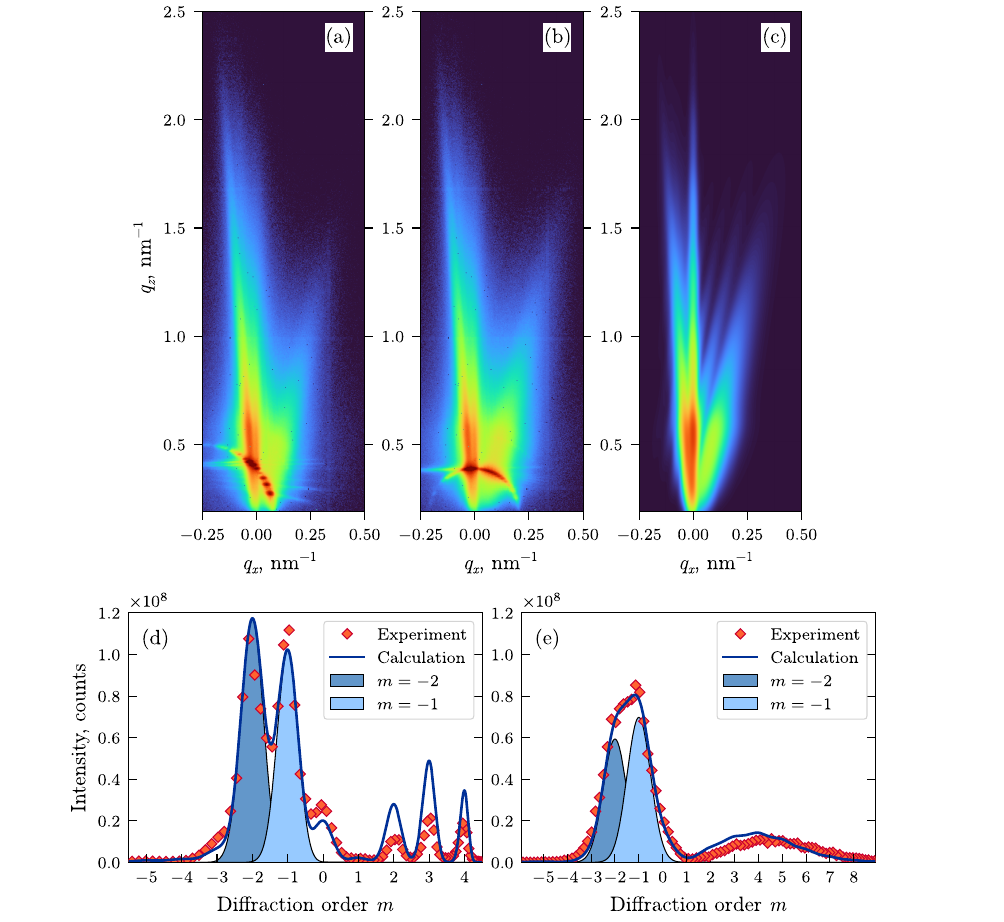}
        \caption{
                    (a,b) GISAXS data for the blazed grating sample, 
                    intensity is shown in logarithmic color-scale.
                    The exposure parameters are $\alpha_i = 0.25^\circ$, $E = 9$~keV.
                    The pattern in (b) is measured at conical diffraction ($\phi_i = 0^\circ$),
                    while the pattern in (a) is measured for the grating rotated laterally by
                    $\phi_i = 0.3^\circ$ degrees.
                    (c) Numerical simulation of diffuse scattering.
                    (d,e) Numerical simulation of diffraction compared to experimental data. Experimental data (markers) are prepared in form of line sections of GISAXS patterns along the diffraction cone.
                    The data in (d) and (e) are for $\phi_i=0.3^\circ$
                    and $\phi_i=0^\circ$ respectively.
                    Theoretical curves (solid lines) are calculated as the sum of the diffraction peaks.
                    Diffraction peaks for $m = -1$ and $m=-2$, are highlighted.
                }
        \label{fig:main}
    \end{figure}

    Measured GISAXS patterns are shown in Fig.~\ref{fig:main}~(a,b).
    Both patterns were measured with an angle of incidence of $\alpha_i = 0.25^\circ$ and a photon energy of $E = 9$~keV.
    The pattern in Fig.~\ref{fig:main}~(b) was measured in the conical diffraction setting,
    i.e., $\phi_i = 0^\circ$,
    while the pattern in Fig.~\ref{fig:main}~(a) was measured for the tilted conical geometry at $\phi_i = 0.3^\circ$.
    This scattering geometry is shown schematically in Fig.~\ref{fig:sketch}~(c).

    Let us first look at the pattern measured at $\phi_i = 0^\circ$ in Fig.~\ref{fig:main}~(b). 
    There are two main features:
    a bright circle around $q_x = q_z = 0$,
    which is associated with conical diffraction;
    and long streaks of diffuse scattering extended in the $q_z$ direction. There are also bright thin streaks extending along $q_x$,
    which we associate with diffraction at the optical system of the beamline itself,
    so we disregard them in the following considerations.
    Notably, the diffraction in this pattern has a form of continuous intensity distribution along the cone section, with no distinction between discrete diffraction orders.
    In fact, for given parameters ($\alpha = 0.25^\circ$, $D/\lambda \sim 2900$)
    according to Eq.~\ref{eq:spherical},
    there should be 21 of diffraction maxima on the cone section and there is not enough instrument resolution to resolve them.
    On the contrary, in the pattern measured at $\phi_i = 0.3^\circ$ [Fig.~\ref{fig:main}~(a)], the diffraction cone section is shifted and the individual peaks are visually distinguishable.
    Meanwhile, the diffuse scattering (streaks along $q_z$) remained unchanged.

    This is more clearly visible in line cuts from the GISAXS data along the diffraction cone.
    The markers in Fig.~\ref{fig:main}~(d,e) represent the line cuts.
    To prepare the line cuts, each pixel of the detector was assigned a coordinate in reciprocal space. Then a histogram was constructed from the data around the diffraction cone.
    In this way, an actual number of photon counts is preserved.
    Finally, the $q_x$ coordinates were mapped to the order of diffraction as $m = q_xD/2\pi$, so that
    each diffraction peak maximum should be observed near an integer of $m$.
    The diffraction peaks in Fig.~\ref{fig:main}~(e) (measured at $\phi_i = 0^\circ$) are not resolved,
    as there are only two broad maxima along the diffraction cone.
    This is in contrast to Fig.~\ref{fig:main}~(d) where the sample is measured at $\phi_i = 0.3^\circ$.
    Indeed, there are fewer diffraction orders distributed along the measured part of the diffraction cone (the density of $m$),
    and diffraction peaks are resolved.

    As a model of the structure for all following calculations we use the average profile of the grating groove shown in Fig.~\ref{fig:AFM} obtained by the AFM. Since the angle of incidence $\alpha_i = 0.25^\circ$ is smaller than the critical angle of platinum: $\alpha_c = 0.52^\circ$ at photon energy $E = 9$~keV, we consider the grating to consist completely of platinum,
    ignoring the silicon base of the structure for the simulations.
    It is negligible since less than $7.5\cdot10^{-5}$ of X-rays penetrate platinum deeper than 25~nm under these illumination conditions.
    Therefore, for the following calculations we will consider the triangular structure taken from real data, the material inside this structure is homogeneous with optical constants:
    $\delta=3.7 \cdot 10^{-5}$ and $\beta = 3.1\cdot10^{-6}$, which are taken from~\cite{Henke93}.

    Mathematically, the problem of conical diffraction is formulated with the 3D Helmholtz equation using rigorous boundary conditions and illumination conditions~\cite{Popov14}, in which the structure of a grating is represented as a potential that is periodic in the lateral direction and the incident plane wave has any direction and polarization. In such a setting, the far-field solution is represented by a series of delta-like diffraction peaks, i.e., the solution is represented by a discrete set of diffraction amplitudes $R_m$ which corresponds to the intensity of diffraction peaks arranged in a cone as $I\propto|R_m|^2$. There are many ways to rigorously compute the set of diffraction amplitudes, such as the surface coordinate transform (C method)~\cite{Chandezon1980}, the finite difference time domain approach~\cite{Popov14}, the finite element method~\cite{Pomplun2007}, the volume integral equation method in curvilinear coordinates~\cite{Shcherbakov2013}, and many others~\cite{Popov14}. In this study, we used the method based on boundary integral equations (BIEM), first proposed for X-ray gratings in~\cite{Goray94} and described in details in~\cite{Popov14}, and the Fourier modal method (FMM)~\cite{Moharam82}, whose modification, which we specifically use, is described in~\cite{Nikolaev24}.
    For the BIEM simulations we used the PCGrate$^\mathrm{TM}$ software package~\cite{Goray05},
    and for the FMM approach we used a special variant of this method that includes elements of dynamic diffraction theory and treats the structure of the grating as a free-form polygon~\cite{Nikolaev24}.
    Thus, in both methods the diffraction amplitudes $R_m$ can be calculated directly from the AFM data.

    \begin{SCfigure}[][t!]
        \centering
        \includegraphics{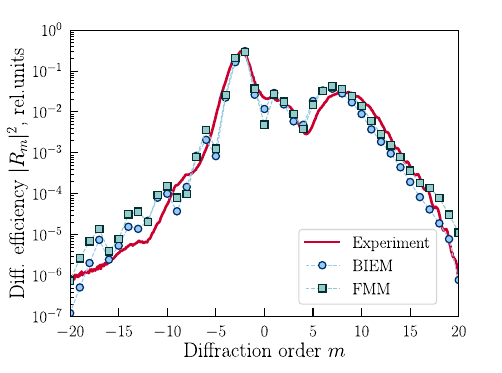}
        \caption{Comparison of two calculation methods.
                 The Boundary Integral Equation method (BIEM, circle markers)
                 and the Unified Fourier Modal method (FMM, square markers).
                 The calculation results are presented as diffraction amplitudes $|R_m|^2$ on a logarithmic scale.
                 The interpolated line cut from experimental data along the diffraction cone shown for the reference,
                 the experimental curve is linearly scaled to fit the simulations.
                 The exposure parameters are
                 $E = 9$~keV, $\alpha_i = 0.4^\circ$, $\phi_i = 0^\circ$.
                }
        \label{fig:comparison}
    \end{SCfigure}

    First, we compared the simulations performed by BIEM and FMM.
    The results are shown in Fig.~\ref{fig:comparison}, where the discrete diffraction amplitudes calculated by both methods are shown as markers. The simulation is performed for conical diffraction $\phi_i = 0^\circ$ and for angle of incidence $\alpha_i = 0.4^\circ$. These results are plotted against the experimental data represented by an interpolated line cut along the diffraction cone. The data and simulations in Fig.~\ref{fig:comparison} are on the logarithmic scale.
    The line cut from the experimental data is interpolated for visual clarity.
    The results are in good qualitative agreement both between BIEM and FMM and between simulation and experiment taking into account the average groove profile measured.

    Next, we simulated the diffraction patterns in Fig.~\ref{fig:main}. Again, in the real experiment, the diffraction pattern is continuous rather than descrete due to the imperfections of the structure and the limited instrumental resolution of the beamline. To account for this in the simulations, we first computed the diffraction amplitudes $R_m$. Then we computed a set of Gaussian profiles, each centered at the respective diffraction order coordinate $m$, and scaled each Gaussian profile by the corresponding diffraction efficiency $|R_m|^2$.
    For reference, we show these separate diffraction peaks for diffraction orders $m=-1$ and $m=-2$ in Fig.~\ref{fig:main}~(d,e).
    The resulting theoretical curve is the sum of all these scaled Gaussians. Finally, the widths of the peaks are adjusted for best fit to the experimental data.
    All the simulations are in good qualitative agreement with the experiment.

    It is clearly visible in Fig.~\ref{fig:main}~(e) that two diffraction peaks ($m=-1$ and $m=-2$) are overlapping in conical diffraction resulting in a single broad peak. Thus, instead of resolving each peak individually, one can see two bright broad peaks in Fig.~\ref{fig:main}~(e). These two peaks are actually attributed to the scattering on the corresponding edges of the blazed grating profile. The peak at $m = -1,-2$ corresponds to the reflection at the blazed edge, and the one at $m = 2,3,4,5$ is due to the anti-blazed edge. Another characteristic feature of the blazed gratings is the suppressed specular reflection, as the peak at $m = 0$ is lower than the blazed peaks at $m=-1,-2$.

    We have demonstrated a qualitative agreement between the simulations and the experimental data. Although, there are visible numerical discrepancies with the experiment. Again, our forward model is based on the AFM pre-characterization (Fig.~\ref{fig:AFM}). This may be the source of the discrepancy, as the AFM characterization is local and the statistics of the studied area may differ from the whole structure which is illuminated in the conical diffraction measurements. However, the pattern of the experimental data is well reproduced.

    The diffuse part of the pattern is due to the scattering on the surface of the grating. 
    The diffuse pattern itself consists of two branches:
    one at $q_x<0$ is tilted to the left and another to the right.
    One can give a simplified qualitative explanation for such a pattern.
    In the case of blazed gratings, there are two sets of surfaces, blaze and anti-blaze. The two branches are oriented in reciprocal space so that they are perpendicular to the corresponding facets. In fact, the left branch is tilted from $q_z$ axis by about
    $4^\circ$, which corresponds to the blaze angle $\gamma_{\rm b}$.
    The left branch also has a structure: it consists of two bright streaks. The diffuse scattering pattern cannot be fully explained by geometrical considerations of the grating, as interference effects must be taken into account.
    To reproduce the diffuse scattering pattern semi-quantitatively, we performed another set of numerical simulations.


    For these simulations we again use the groove profile from AFM measurements~(Fig.~\ref{fig:AFM}).
    We use the distorted wave Born approximation (DWBA) theory~\cite{Sinha88,Kaganer95} to calculate the amplitude of the scattered wave.
    The DWBA is a perturbation theory. It is implemented by separating the structure 
    onto the ideal part, for which the wave equation can be solved exactly, and the perturbation part.
    This partitioning can be done in several ways.
    For simplicity, we perform this partitioning in a manner similar to that used in these works~\cite{Mikulik98}.
    Namely, the substrate is considered as an ideal part of the structure.
    In this way, the solution of the wave equation can be easily written as a standing wave with Fresnel coefficients as its amplitudes.
    Next, we consider the grating itself as a perturbation on top of the substrate. We further simplify the model by considering the grating not as a periodic array of triangle lines,
    but as a single line on the surface. Considering the periodicity, one would take into account conical diffraction in the sense of kinematical scattering.
    Such an approximation is not just for the case of conical diffraction, moreover, we have already calculated the diffraction in the previous simulations with several rigorous approaches.
    By considering only a single line of the grating, we assume that photons scattered on surfaces from two different lines are not correlated in phase and contribute to the scattering pattern incoherently, i.e. by simply scaling the scattered intensity.
    The resulting equation for the diffuse scattering amplitude is
    $f(\bm{q}) \propto t_f \hat V(\bm q)$,
    where $t_f$ is the Fresnel transmittance of the substrate, calculated for the scattering direction $\bm{k}_f$,
    and $\hat V(\bm q)$ is the form factor,
    i.e. the Fourier transform of a function $V(\bm{r})$
    that is $V=1$ for $\bm r$ inside the body of the line (colored area in Fig.~\ref{fig:AFM})
    and $V=0$ otherwise.
    For computational efficiency, this can be done by integrating over the line segments of a groove profile in Fig.~\ref{fig:AFM},
    as described, for instance, in~\cite{Kaganer21}.
    As such, this approach takes into account the fact that diffuse scattering is proportional to the gradient of electron density, i.e.,
    it occurs at the surface of the sample.
    However, it does not take into account the morphology of the surface roughness.
    The calculation is based on an empirical approach.
    We considered that there should be a decay in intensity with respect to $q$,
    which effectively accounts for a finite correlation length of the roughness along the surface. Thus, we considered the diffuse scattering intensity as 
    $
        I \propto|t(q_z)\hat{V}(\bm{q})|^2 \eta(\bm{q}),
    $
    and the result is shown in Fig.~\ref{fig:main}~(c).
    The term $\eta$ is a Debye-Waller term:~$\eta(\bm{q}) = \exp(-q_x^2s_x^2) \exp(-q_y^2s_y^2) \exp(-q_z^2s_z^2)$;
    with the coefficients $s_{x,y,z}$, which are arbitrarily chosen to qualitatively match the diffuse scattering pattern in the experimental data.
    
    In fact, the simulated diffuse scattering pattern [Fig.~\ref{fig:main}~(c)]
    is in excellent qualitative agreement with the experimental data  [Fig.~\ref{fig:main}~(a,b)].
    One can see the long streaks extending along the $q_z$ direction.
    There are two branches of these streaks.
    The first branch is a single wide streak in the area of $q_x>0$ and the second branch consists of two narrow bright streaks in $q_x<0$.
    Notably, these two streaks are crossing the origin of the reciprocal space and the position corresponding to the $m=-1$ and $m=-2$ orders of diffraction.
    This is observed in the data as well as in the simulation.
    Therefore, these streaks can be attributed to diffuse scattering of the diffracted beams. Indeed, there are two diffraction orders whose diffraction efficiency is enhanced by the blaze ($m= -1,-2$),
    and the bright streaks correspond to the same orders of diffraction.
    Remarkably, diffraction and diffuse scattering have been calculated using the same model,
    but approaches based on completely different theoretical concepts. Nevertheless, the intensity enhancement at $m=-1,-2$ is reproduced in both cases. Thus, in this case we can interpret the diffuse scattering as a "leakage" of the diffraction efficiency due to the roughness of the reflection surface.
    This underscores the usefulness of synchrotron measurements, as the effect of structural imperfections on optical performance is directly observed.
    Moreover, in the GISAXS experiment, one can not only see the loss of diffraction efficiency, but also pinpoint where this lost fraction of photons has gone.

    This immediately raises the problem of quantitatively characterizing the diffuse scattering of radiation diffracted on a blazed grating.
    Within the DWBA,
    this can be done rigorously,
    however only for the planar systems such as thin films.
    For a simple structure, such as a planar surface,
    the asymptotics of diffuse scattering can be interpreted as the power spectral density (PSD)~\cite{pelliccione2008} of the surface roughness.
    This feature of diffuse scattering has been exploited in works~\cite{Stone99,Chkhalo14}.
    Particularly in~\cite{Chkhalo14}, the scattering distributions have been compared with the PSD estimates from AFM data.
    The scattering data and the AFM estimates overlap, but cover different ranges of spatial frequencies.
    The curves are in agreement in the overlapping region.
    Thus, the AFM and scattering measurements are shown to be complementary.
    For more complex structures, such as thin films with correlated roughness at different interlayers,
    the scattering pattern cannot be directly interpreted as a PSD.
    An elegant interpretation has been proposed in~\cite{Siffalovic09}.
    The idea is to consider a certain direction in reciprocal space for which the diffuse scattering is enhanced due to interference,
    and to examine the intensity distribution along this direction.
    In~\cite{Siffalovic09} the diffuse scattering on the periodic multilayer was considered and the sections of the Bragg sheets along the $q_x$ direction were found to be proportional to the PSD of the roughness averaged across the buried interfaces.

    \begin{SCfigure}[][t!]
        \centering
        \includegraphics{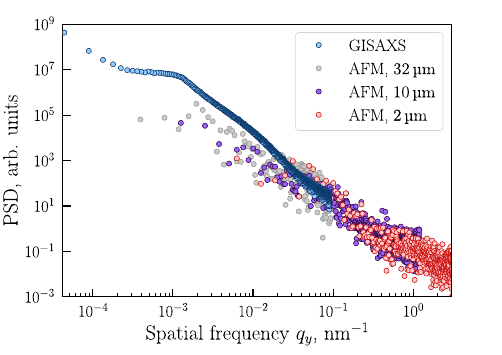}
        \caption{
                Power spectral density function as estimated from the GISAXS data (blue markers) and from the AFM data measured
                at \SI{32}{\micro\metre} scale (gray markers),
                at \SI{10}{\micro\metre} scale (violet markers),
                and at \SI{2}{\micro\metre} scale (red markers).
                The values of PSD are given in arbitrary units scaled to the integrated photon count of the GISAXS data.
        }
        \label{fig:PSD}
    \end{SCfigure}

    To the best of our knowledge, the problem of calculating scattering on the nanopatterned surfaces has not been rigorously solved.
    Similarly to~\cite{Siffalovic09}, we propose simply choosing the direction in which scattering is enhanced by interference and examining the distribution of scattering intensity along that direction.
    In our case, scattering enhancement corresponds to the blaze angle.
    Namely, we consider the line cut along the bright streak corresponding to the diffraction order $m=-2$ [cf. Fig~\ref{fig:main}(b)].
    Note that the pattern in the figure is plotted according to the $q_x$ and $q_z$ coordinates.
    However, each pixel on the detector corresponds to a specific point in reciprocal space, which has three coordinates: $q_x$, $q_y$ and $q_z$.
    The $q_y$ coordinate corresponds to the momentum transfer along the grooves,
    therefore $q_y$ can be interpreted as the spatial frequency in that direction.
    The line cut taken from the GISAXS data in Fig.~\ref{fig:main}~(b) for the $m=-2$ intensity streak is plotted against the corresponding $q_y$ coordinate in Fig.~\ref{fig:PSD} in a logarithmic scale
    (blue markers).
    We hypothesize that this curve is proportional to the power spectral density (PSD) of reflective surface roughness.
    
    Following this hypothesis,
    the scattering line cut curve can be interpreted as follows. 
    First, there is the $m=-2$ diffraction peak itself at the beginning of the curve.
    Once again, it corresponds to the groove profile rather than the roughness.
    Next, there is a plateau attributed to a corresponding spatial correlation scale: $\xi = \SI{0.8}{\micro\metre}$.
    Further, the curve has a downward tail.
    Its tangent $k$ corresponds to the Hurst $H$ parameter, which, in turn, is connected to the fractal dimension of the surface roughness.
    For a two-dimensional surface, $k = 2H + 1$.
    As shown in Fig.~\ref{fig:PSD},
    the tangent of a tail changes slightly at higher spatial frequencies.
    The estimated value of the Hurst parameter 
    based on an averaged slope of the tail is $H=0.33$.
    
    To test whether the line cut from the GISAXS corresponds to the PSD of the reflective surface roughness,
    we compared it to the PSD estimated from the AFM measurements.
    For this estimate, we naturally used the AFM data obtained after the reflective coating was deposited.
    The PSD is estimated as the absolute value squared of a Fourier transform of the AFM profiles.
    To extend the frequency-limited range, three measurements were taken at different spatial scales. These scales are indicated in the legend of Fig.~\ref{fig:PSD}.
    The \SI{2}{\micro\metre} scale data was truncated at a spatial frequency of $q_y = 5$~nm$^{-1}$ because the curve contained features associated with analog-to-digital converter artifacts.
    The PSD curve is stitched together from three overlapping curves and scaled to the GISAXS line cut,
    which uses photon count units.
    There is a smooth transition from the synchrotron data to the AFM data at $q_y~\sim[10^{-2},\, 10^{-1}]$~nm$^{-1}$. This suggests that the two methods agree for roughness morphology at a spatial scale of 1~nm to 50~nm.
    There is a slight discrepancy in $q_y~\sim[10^{-3},\, 10^{-2}]$~nm$^{-1}$ range. However, the AFM curve is limited by resolution and noise, which maybe the reason for discrepancy. 
    Thus far, the agreement at
    $q_y > 10^{-2}$~nm$^{-1}$
    allows us to assume that, in blazed grazing, one can analyze the PSD of the roughness selectively on the reflective surface.
    In this work, we demonstrated that synchrotron measurements on blazed gratings are highly sensitive to the groove profile details and reflective surface roughness statistics.

\section{Conclusions}

    We examined the utility of synchrotron measurements as a metrology tool for blazed gratings.
    For the test sample, we used a blazed grating with a 400~nm period.
    The groove pattern was printed on a silicon substrate via the anisotropic etching through an e-beam lithography mask.
    The grating was then coated with a reflective platinum film.
    This grating is specifically designed for the use in the X-ray spectral range in the grazing-incidence conical mountings.
    The GISAXS measurements on the test sample were performed at a bending magnet synchrotron source.
    In the experiment, both the conical diffraction pattern and the diffuse scattering were observed.

    The interest in such synchrotron measurements is twofold.
    First, the performance of the blazed grating as an optical element can be directly observed in this experiment.
    Namely, it allows for the measurement of the diffraction efficiency of the grating,
    as well as for direct measurement of loss of efficiency into diffuse scattering.
    The absolute diffraction efficiency of the test sample, which was measured and rigorously calculated using the groove profile from AFM,
    is approximately 30\% in the $m = -2$ order
    at a wavelength of $\lambda = 1.38 $~\AA.
    One highlight is that the optical device, which has a sub-micron characteristic size, can operate at an \AA{}ngstrom-level wavelength due to the high dispersion in the conical mounting.
    That said, the resolution was insufficient to distinguish the diffraction peaks at the exact conical mount ($\phi_i = 0^\circ{}$).
    Nevertheless, we avoided this issue by tilting the sample at a small angle of $\phi_i = 0.3^\circ$ around the vertical axis.
    Moreover, we demonstrate that it is possible to resolve at least three distinct diffraction peaks
    ($m = -1, 0$~and~$1$) for any wavelength-to-period ratio. This is again due to the spherical dispersion.
    
    The second utility for such measurements is to reconstruct the groove profile and evaluate the roughness of the reflective surface.
    This research did not involve model reconstruction from GISAXS data. Instead, we ran forward numerical simulations of scattering and diffraction based on a model pre-characterized by an AFM.
    This is done to demonstrate that the model-based simulation reproduces the observed features,
    thereby showing that the GISAXS data encodes relevant information.
    Firstly, we considered the diffraction efficiency.
    For these simulations, we used two rigorous solvers for the Helmholtz equation: FMM and BIEM. For the forward model, we used the groove profile from the AFM measurements. We compared the diffraction simulation results between BIEM and FMM, and then compared them with the experimental data.
    All of these comparisons demonstrate good quantitative agreement.
    Secondly, we examined the diffuse scattering.
    Using a simple model based on the DWBA,
    we were able to reproduce the diffuse scattering pattern.
    Furthermore, by comparing this analysis to the data, we showed that one can identify the specific part of the pattern corresponding to scattering on a rough, reflective surface.
    We extracted a PSD function for the roughness of the reflective surface by taking a line-cut from the data in this direction.
    Then, we compared the PSD extracted directly from the scattering data with the PSD calculated from the AFM.
    The curves agree well within the spatial frequency range
    that corresponds to the scale of 1~nm to 50~nm.
    The sensitivity and selectivity of the scattering method with respect to the roughness of the reflective surface of the blazed grating has yet to be proven. Nevertheless, this method evidently provides valuable information on the roughness morphology.

    Thus, one can collect information on the groove profile and reflective surface morphology of a blazed grating and measure its efficiency in a single exposure at a synchrotron source.
    Taken together, these results suggest that synchrotron measurements could be extremely useful for designing and manufacturing such optical elements.
    With this, we hope to once again draw the attention of the diffractive optics community to the GISAXS as a nanometrology tool,
    as it can help with future advances in EUV and X-ray optics.
    
\section*{Acknowledgements}

    The authors acknowledge the scientific infrastructure of the Kurchatov Synchrotron Radiation Source "KISI" at the NRC "Kurchatov Institute", which made the synchrotron measurements possible.
    One of us (KN) owes a debt of gratitude to 
    Dr.~Anal\'{i}a Fern\'{a}ndez Herrero
    and
    Dr.~Victor Soltwisch for sharing their expertise and advice.

\section*{Funding information}

    Part of the work related to sample design and fabrication is supported by the Russian Science Foundation (the grant number 25-12-00139). Part of the work related to synchrotron measurements, data analysis and scattering simulations is supported by the Russian Ministry of Education and Science (the contract number 075-15-2025-456 signed on May 30, 2025).

\bibliographystyle{ieeetr}
\bibliography{bibliography}    
\end{document}